\documentclass[a4paper,11pt]{article}
\usepackage{jinstpub} 
\usepackage{lineno}

\usepackage{comment,mfirstuc}
\usepackage{enumitem}
\usepackage{hyperref}
\usepackage{graphicx}
\usepackage{caption}
\usepackage{subcaption}
\usepackage{multirow}
\usepackage{tabularx}
\usepackage{xspace}

\usepackage{hyperref}

\usepackage{tikz}
\usetikzlibrary{shapes,arrows,positioning}

\newcommand{\addComment}[2]{
  \expandafter\newcommand\csname #1\endcsname[1]{{\bf \color{#2} \capitalisewords{#1}:\,##1}}
  \expandafter\newcommand\csname #1cor\endcsname[2]{{\color{#2} \capitalisewords{#1}:\,\st{##1}{\bf ##2}}}
  \expandafter\newcommand\csname #1color\endcsname{#2}
}
\addComment{anselm}{magenta}
\addComment{cris}{blue} 
\addComment{hemalata}{blue} 
\addComment{karthik}{red}

\newcommand{\geant}{\textsc{Geant4}\xspace}
\newcommand{\epic}{\textup{ePIC}\xspace}

\title{AI-Assisted Detector Design for the EIC (AID(2)E)}

\author[4]{M. Diefenthaler}
\author[5]{C. Fanelli}
\author[3]{L. O. Gerlach}
\author[1]{W. Guan}
\author[2]{T. Horn}
\author[1]{A. Jentsch}
\author[1]{M. Lin}
\author[3]{K. Nagai}
\author[5]{H. Nayak}
\author[3]{C. Pecar}
\author[5]{K. Suresh}
\author[3,4]{A. Vossen}
\author[1]{T. Wang}
\author[1]{T. Wenaus}
\author[]{(AID(2)E collaboration)}

\affiliation[1]{Brookhaven National Laboratory}
\affiliation[2]{Catholic University of America}
\affiliation[3]{Duke University}
\affiliation[4]{Thomas Jefferson National Laboratory}
\affiliation[5]{William and Mary}

\emailAdd{cfanelli@wm.edu}

\abstract{
Artificial Intelligence is poised to transform the design of complex, large-scale detectors like \epic at the future Electron Ion Collider. Featuring a central detector with additional detecting systems in the far forward and far backward regions, the \epic experiment incorporates numerous design parameters and objectives, including performance, physics reach, and cost, constrained by mechanical and geometric limits.
This project aims to develop a scalable, distributed AI-assisted detector design for the EIC (AID(2)E), employing state-of-the-art multiobjective optimization to tackle complex designs. Supported by the \epic software stack and using \geant simulations, our approach benefits from transparent parameterization and advanced AI features.
The workflow leverages the PanDA and iDDS systems, used in major experiments such as ATLAS at CERN LHC, the Rubin Observatory, and sPHENIX at RHIC, to manage the compute intensive demands of \epic detector simulations. Tailored enhancements to the PanDA system focus on usability, scalability, automation, and monitoring.
Ultimately, this project aims to establish a robust design capability, apply a distributed AI-assisted workflow to the \epic detector, and extend its applications to the design of the second detector (Detector-2) in the EIC, as well as to calibration and alignment tasks. Additionally, we are developing advanced data science tools to efficiently navigate the complex, multidimensional trade-offs identified through this optimization process.
}

\keywords{{\color{blue}Artificial Intelligence}, {\color{blue}Detector Design}, {\color{blue}Electron Ion Collider}, {\color{blue}Multi-Objective Optimization}}

\begin{document}
\maketitle
\flushbottom

\section{Introduction}
\label{sec:intro}


The Electron Ion Collider (EIC) stands out as one of the pioneering large-scale experiments to incorporate artificial intelligence (AI) right from its design phase. As highlighted in the 2019 DOE Town Halls on AI for Science: ``\textit{AI techniques that can optimize the design of complex, large-scale experiments have the potential to revolutionize the way experimental nuclear physics is currently done}.''
This paper presents AID(2)E, a DOE-funded initiative aimed at developing a scalable and distributed framework for AI-assisted detector design at the EIC. 

The project was initiated to address the limitations of current detector optimization methods. Traditionally, each subsystem is optimized individually within global design constraints before integration. This conventional approach, often relying on ad hoc or brute-force methods like grid search, is sub-optimal. It struggles with handling multiple design parameters, fails to optimize multiple objectives simultaneously (such as efficiency, resolution, and costs), and lacks holistic optimization of the entire detector. Furthermore, it does not fully exploit the benefits of multi-objective optimization, which can reveal optimal trade-offs between competing objectives.
The rationale for employing a multi-objective optimization approach in detector design is that there is no single optimal solution; instead, there exists a spectrum of viable solutions that balance factors such as detector response, physics design requirements, and costs. This is especially significant given the high costs associated with large-scale detectors, exemplified by the estimated \$300M budget for the ePIC detector \cite{ent_nsac_lrp}.
As demonstrated in previous studies (e.g., \cite{Cisbani-2020, Fanelli:2022nima}), these methods provide new insights into internal correlations among design parameters and enable trade-offs that are not achievable with traditional methods. AI-assisted optimization complements the expertise of detector specialists and, as shown in \cite{Fanelli:2022nima}, can improve objectives compared to a baseline design across a large phase space.
A holistic optimization of the entire detector, considering multiple objectives simultaneously, represents a paradigm shift in detector design. AI-assisted methods can significantly impact future large nuclear physics projects, such as those at the EIC. Using realistic simulation pipelines and minimizing the computational budget required, the AID(2)E Collaboration aims to streamline the design process while ensuring high-fidelity results. The primary goal of AID(2)E is to develop infrastructure that optimizes detector design in a scalable and distributed manner. The project begins with two closure tests designed to demonstrate the convergence of the Pareto front in multi-objective optimization problems and to validate the effectiveness of distributing the compute-intensive simulations of design points suggested by the optimizer.
AID(2)E has already made contributions towards optimizing the ePIC detector subsystems. Currently, the collaboration is optimizing two detector subsystems, which are discussed in this proceeding. This optimization approach can potentially be extended to other subsystems of ePIC, making it one of the first large-scale detector designed with the assistance of AI.
Evaluating different solutions on the Pareto front of ePIC at various budget levels can yield significant cost benefits. Even fractional improvements in objectives, such as resolution and efficiency, can lead to a more efficient use of beam time, a major cost factor for the EIC over its lifetime. Characterizing the Pareto front can inform the design of a complementary Detector-2 at EIC. Our team is also developing a suite of data science tools to facilitate effective exploration of the determined Pareto front \cite{pareto_interactive}. It is worth mentioning that the AID(2)E framework also holds potential for other applications, such as calibration and distributed machine learning inference engines. Furthermore, AID(2)E aims to engage the community through bootcamps and short-term schools for students.

\section{Methodology and Strategy}
\label{sec:mthodology}

The primary goal of the AID(2)E project is to develop a scalable and distributed AI-assisted Detector Design framework for the EIC, specifically focusing on the \epic detector. Central to AID(2)E is Multi-Objective Optimization (MOO), which evaluates various design objectives such as sub-detector intrinsic responses (\textit{e.g.}, resolution and efficiency), performance for key EIC physics channels, and material cost proxies.
%
This project is pioneering in its application of MOO at such a scale, facilitated by a developing distributed workflow infrastructure, enhanced with PanDA for simulation workflows \cite{bernauer2022scientific}.
Our strategy includes two closure tests: first, we want to test and characterize the perforamnce of approaches like Multi-Objective Bayesian Optimization (MOBO) \cite{frazier2018tutorial} and Multi-Objective Genetic Algorithms (MOGA), specifically NSGA-II \cite{deb2002fast}, using test functions. The second one proves the effectiveness of distributing the simulations of different design points, coordinating data collection for objective updates, and refining the design iteratively.
%
The framework supports continuous and distributed AI-driven optimization. Initial efforts are concentrated on optimizing specific sub-detector systems within \epic, paving the way for broader application. Detailed discussions on the optimization approach, distributed workflow, and specific applications to the \epic detector are presented in what follows.

\begin{figure}[!ht] 
    \centering
    \includegraphics[trim={0 0 0 0cm},clip,width=0.495\textwidth]{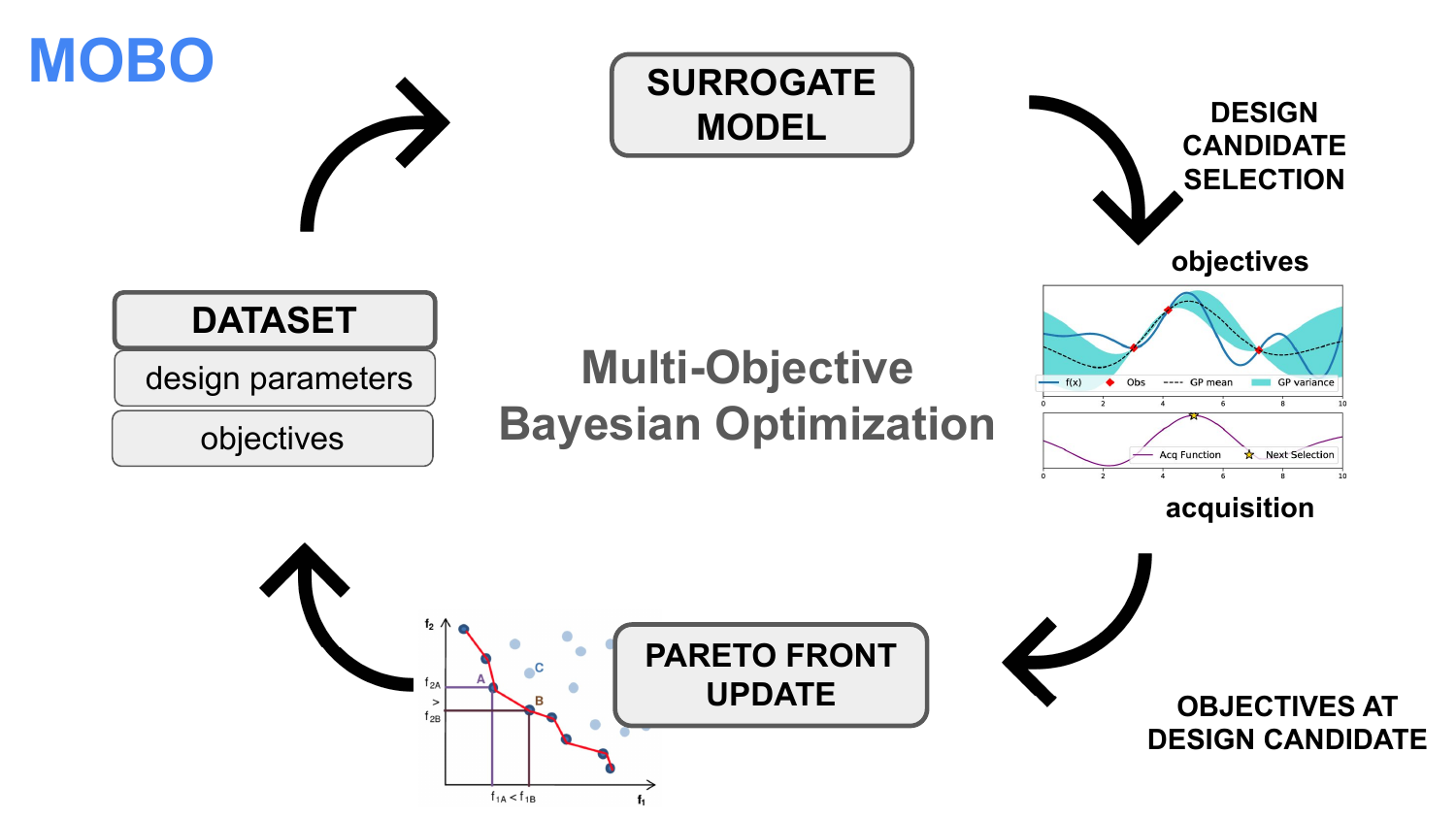}
    \includegraphics[trim={0 0 0 0cm},clip,width=0.495\textwidth]{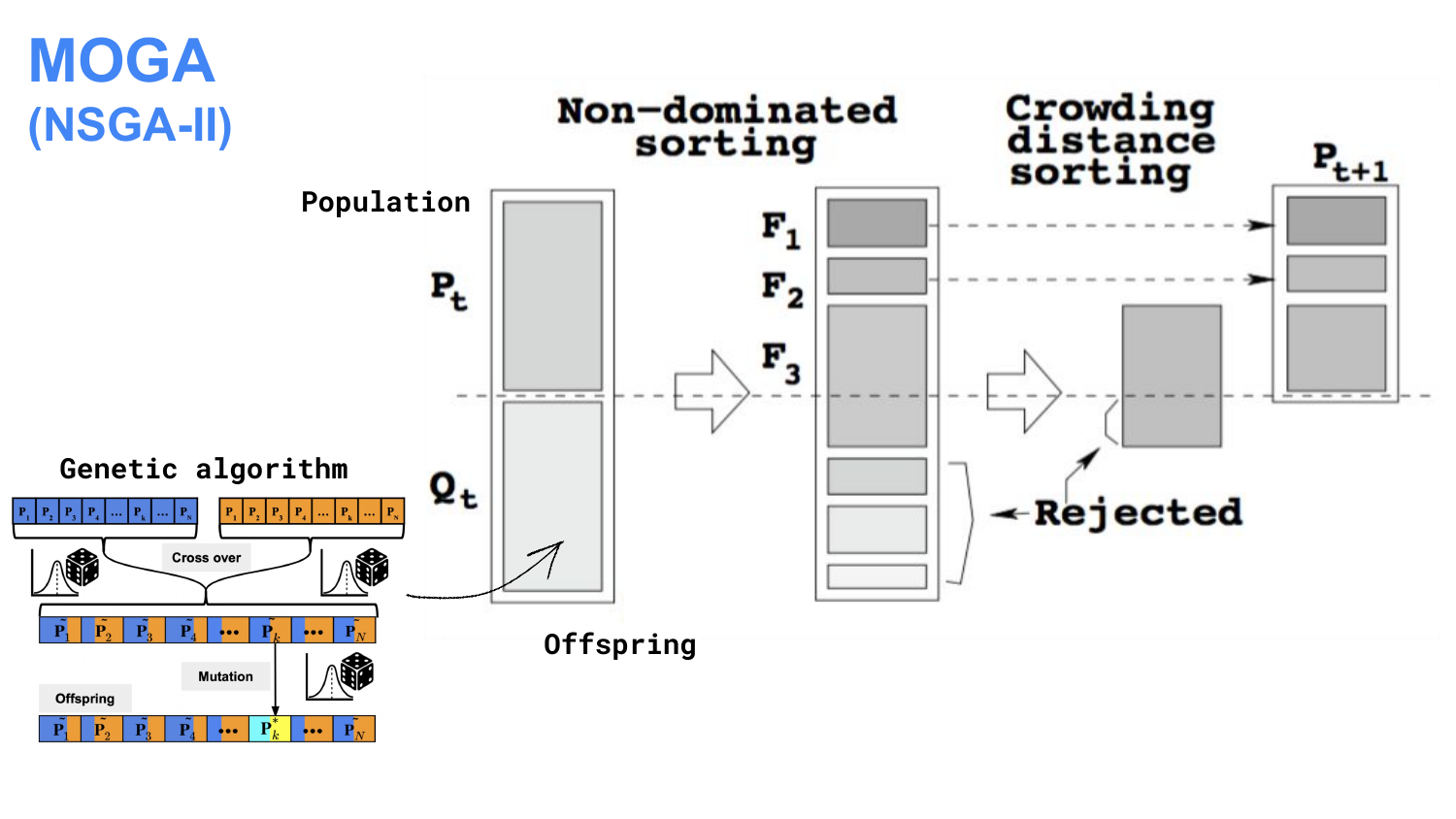}
    \caption{\textbf{Multi-objective optimization workflows}: The left panel shows the Bayesian workflow MOBO, which is the primary choice in this project (generally capable of providing accurate determination Pareto fronts). The right panel shows MOGA (NSGA-II, based on genetic algorithms \cite{deb2002fast}), which is an alternative approach (more suitable to handle, \textit{e.g.}, possible discontinuities or high-dimensional spaces). 
    \label{fig:MOO_workflow}
    }
\end{figure}

\paragraph{Multi-Objective Optimization}
The definition of a generic detector MOO problem can be formulated as follows:
\begin{equation}\label{eq:moo}
\begin{aligned}
        min \ \mathbf{f_{m}}(\mathbf{x})  & \ \ \ \ \ \ m = 1, \cdots, M \ \ \ \mathbf{(objectives)} \\
       s.t. \ \ \ \mathbf{g_{j}}(\mathbf{x})\leq 0, & \ \ \ \ \ \ j = 1, \cdots, J \ \ \ \ \ \mathbf{(inequality \ constraints)} \\ 
       \ \ \ \ \  \mathbf{h_{k}}(\mathbf{x})=0,     & \ \ \ \ \ \ k = 1, \cdots, K \ \ \ \  \mathbf{(equality \ constraints )} \\
       x_{i}^{L} \leq x_{i} \leq x_{i}^{U},         & \ \ \ \ \ \ i = 1, \cdots, N \ \ \ \ \ \mathbf{(parameters)} \\
\end{aligned}
\end{equation}
 where one has $M$ objective functions $f_{m}$ to optimize (\textit{e.g.}, detector resolution, efficiency, signal/background for a physics channel, costs), subject to $J$ inequalities $g_{j}(x)$ and $K$ equality constraints $h_{k}(z)$ (\textit{e.g.}, mechanical constraints), in a design space of $N$ dimensions (\textit{e.g.}, geometry parameters that change the \geant \ design) with lower ($x^{L}$) and upper  ($x^{U}$) bounds on each dimension.
We are utilizing Multi-Objective Bayesian Optimization (MOBO) and Multi-Objective Optimization using Genetic Algorithms (MOGA) to tackle detector design optimization. These techniques assess multiple objectives such as sub-detector responses, physics channel performance, and material costs, and are detailed in our workflow illustrated in Fig. \ref{fig:MOO_workflow}.

MOBO is our primary choice for the \epic detector, driven by our improved understanding of technology and parameter ranges, the ability to incorporate uncertainties and physics-inspired acquisition functions, and a more accurate approximation of the Pareto front compared to MOGA. This approach has proven efficient in requiring fewer evaluations to converge and allows for custom acquisition functions, supported by tools like \texttt{ax-platform} and \texttt{BoTorch} \cite{balandat2020botorch, bakshyopen}. We are currently testing MOBO with a \texttt{qNEHVI} acquisition function \cite{daulton2021parallel}, optimizing for scalability across multiple design points and dimensions.
Scalability challenges for MOBO include computational constraints as the number of observations grows, which we can address using  techniques to manage the increasing complexity of the surrogate model, such as SAASBO and TURBO.
Both MOBO and MOGA offer distinct computational and scalability properties that dictate their application and workload distribution, crucial for the efficient design of the \epic detector. 
One advantage of MOGA compared to MOBO is the possibility of handling nondifferentiable terms, and higher dimensionality problems, leveraging the inherent parallelization of genetic algorithms.

\paragraph{Distributed Workflow}
The AID2E team is committed to advancing AI-based detector design optimization at the EIC by developing a scalable and distributed framework. This framework is refined through its application to the ePIC detector design and is delivered as a robust, experiment-agnostic solution ready for broader detector design applications. The project leverages existing, scalable distributed workflow and workload management tools like PanDA \cite{PanDA-CSBS}, developed for the ATLAS experiment and now supporting complex AI/ML workflows including Bayesian Optimization.

Historically, these systems have supported large-scale data handling and processing needs of the ATLAS experiment and other projects like the Rubin Observatory and sPHENIX, demonstrating the versatility and scalability of the PanDA and the intelligent Data Delivery Service (iDDS). These tools have been integral in managing workflows across diverse computing environments—grids, clouds, and supercomputers—enhancing their stability and adaptability.

Recent developments have focused on integrating complex AI/ML workflows, particularly those utilizing GPUs, into this system. This includes hyperparameter optimization and active learning based on Bayesian optimization, extending the capabilities of ATLAS analyses. Additionally, full-chain integrated production-analysis workflows, including the use of REANA in the analysis phase, have been supported.
The PanDA/iDDS development program emphasizes ease of use and accessibility, supporting a range of AI/ML applications through a unified platform that integrates multiple computing facilities. This capability is critical for processing-intensive, distributed AI-based detector design, offering: (i) seamless operation across distributed resources; (ii) support for comprehensive AI/ML tools and full-chain workflows from simulation to iterative refinement; (iii) efficient authentication and user-friendly interfaces, including support for modern and legacy security mechanisms.
Ultimately, this framework not only supports current project needs but is also adaptable to similar future applications, demonstrating its potential through an implementation of Bayesian Optimization-based active learning now enhancing the physics reach of ATLAS analyses.

\begin{figure}[!ht] 
    \centering
    \includegraphics[trim={0cm 5cm 0cm 0cm},clip,width=0.95\textwidth]{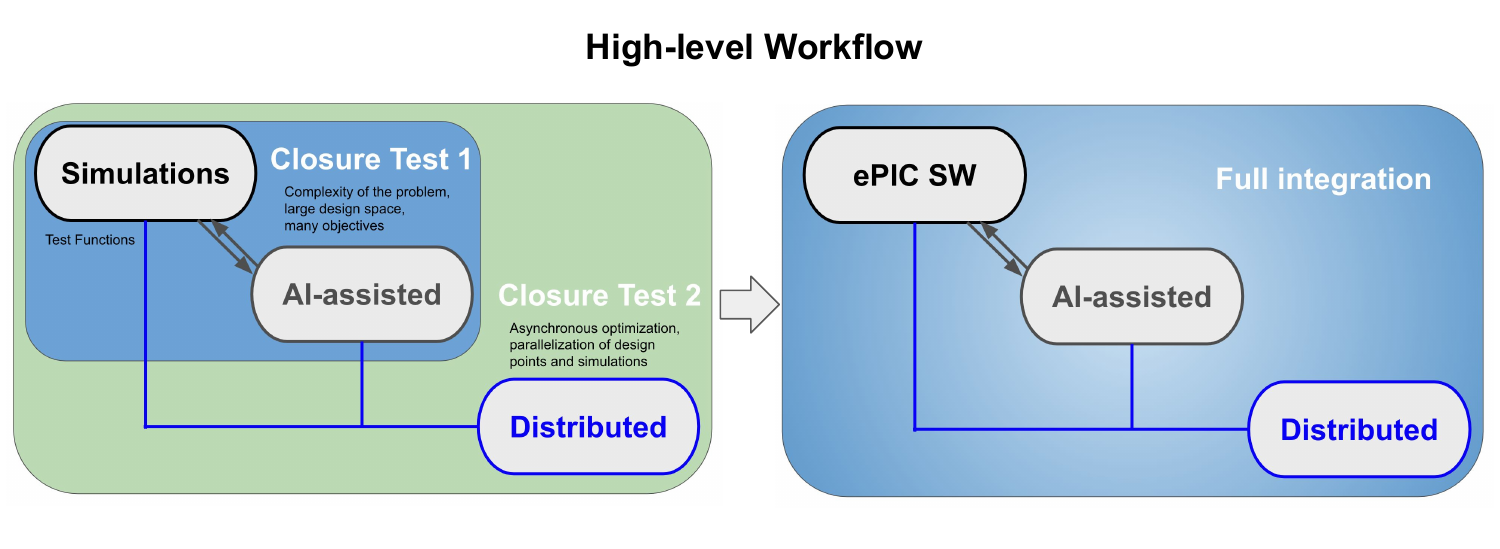}
    \caption{\textbf{High-level workflow of AID(2)E:} We first demonstrate an AI-assisted framework handling multiple test objectives (``simulations'') with a known Pareto front (\textbf{closure test 1}). We also show the feasibility of distributing these simulations across various HPC/HTC clusters using the PanDA framework \cite{PanDA-CSBS} (\textbf{closure test 2}). Following closure test 1, we replace the test functions with the ePIC software (compute-intensive simulations and reconstruction). The final step is a \textbf{full integration}, where the ePIC software is coupled with a distributed AI-assisted framework that asynchronously orchestrates the design points suggested by AI.     
    \label{fig:proposal_workflow}
    }
\end{figure}

\paragraph{Closure Tests}
The primary objective of the closure tests within the AID(2)E framework is to rigorously evaluate system constraints and identify potential bottlenecks. This framework enhances the distribution of simulations using the ePIC software stack \cite{ePIC-software-stack} and advances MOO techniques for detector design. 
The first closure test analyzes these bottlenecks within a test optimization problem, employing MOBO to assess its scalability and efficiency relative to problem complexity.
To meet the computational demands of evaluating multiple detector design points, the \epic detector design optimization requires a large volumes of simulated data. 
The PanDA System \cite{PanDA-CSBS} is utilized to manage these tasks across varied geographic locations, coordinated by the iDDS. The second closure test integrates and stress-tests the PanDA framework, initially with test functions and subsequently with actual \epic simulation payloads. This process validates the framework's capability to efficiently distribute simulations and synchronize them with an optimization strategy, ensuring robust performance across diverse operational scenarios.
A high-level workflow of the AID(2)E framework, based on the mentioned closure tests, is displayed in Fig. \ref{fig:proposal_workflow}.

\begin{figure}[!]
    \centering
    \includegraphics[scale = 0.33]{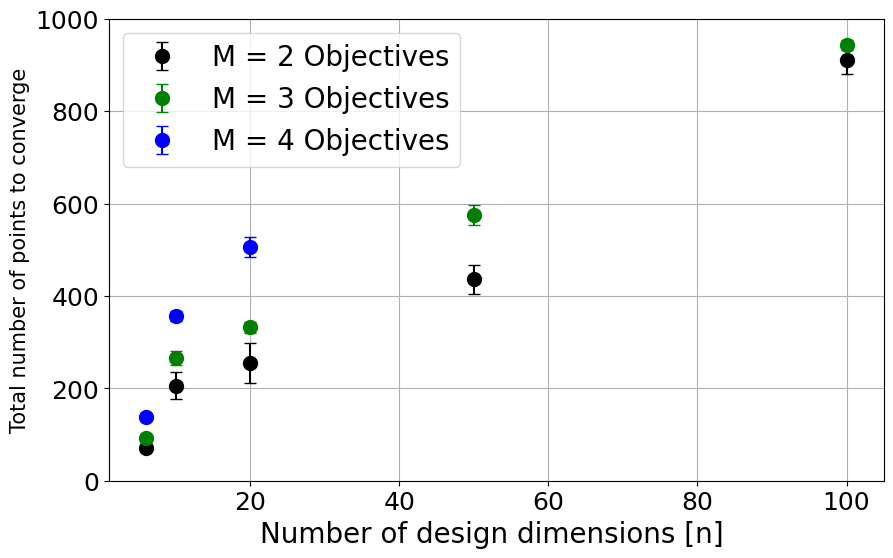}
    \includegraphics[scale = 0.23]{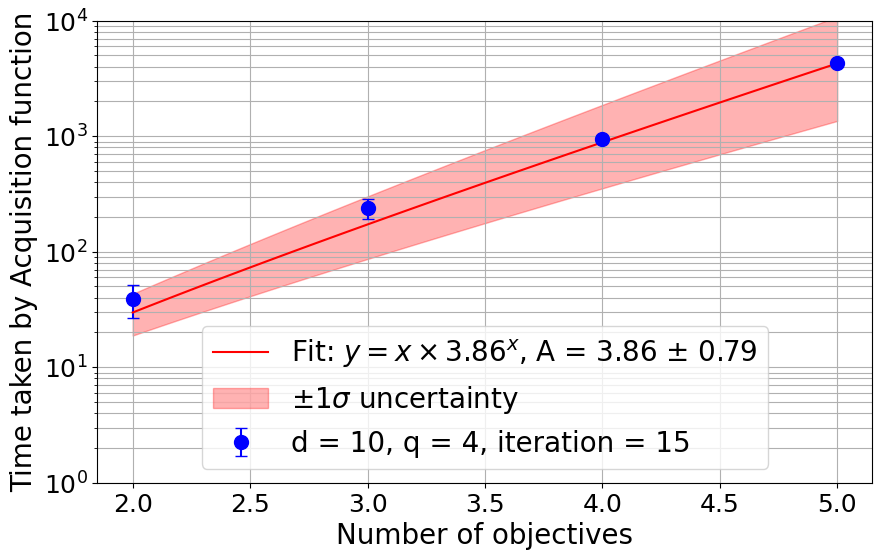}
    \caption{\textbf{Results from closure test}  
    (Left) The figure depicts the total explored points before DTLZ-2 convergence as a function of design parameters ($n$). Each trial was repeated for statistical robustness, with varying colors indicating a different number of objective. The process continued until either 1000 points were explored or convergence was reached, except for cases with $n=[50, 100]$ and $m=4$ where convergence was not achieved. A comparison with expected computational complexity is underway.
    (Right) The figure illustrates the time complexity of the qNEHVI acquisition function. According to \cite{daulton2021parallel}, it is $\mathcal{O}(m\cdot(N+I)^m)$, where $N$ is the observed points and $I$ is partitions in the cached box decomposition. Further studies on varying design parameters and standard latency are in progress to elaborate on the problem's complexity.
    }
    \label{fig:closure-test1}
\end{figure}

%
The primary goal of Closure Test 1 is to evaluate the complexity of the optimization algorithm for convergence with up to 5 objectives and $\mathcal{O}(100)$ design parameters. This test identifies bottlenecks in multi-objective Bayesian optimization algorithms and helps deploy the AID(2)E framework for complex optimization problems. A realistic test function is chosen to pinpoint choke points. Improvements are achieved by reducing computational time and space complexity or the number of iterations needed for convergence. Closure Test 1 uses a test function with a known Pareto front to evaluate convergence. The DTLZ-2 problem is selected for this purpose, defined as follows.
Given $n$ decision variables and $M$ objectives, where $M \geq 2$, the decision vector $x = [x_1, x_2, \ldots, x_n]$ is subject to the constraints $0 \leq x_i \leq 1$ for $i = 1, 2, \ldots, n$, and $g(x) = \sum_{i=M}^{n} (x_i - 0.5)^2 - \cos(20\pi(x_i - 0.5)) \leq 0$. 
The objective functions are defined as:

\begin{align*}
f_1(x) &= (1 + g(x)) \cos\left(x_1 \frac{\pi}{2}\right) \cos\left(x_2 \frac{\pi}{2}\right) \dotsm \cos\left(x_{M-1} \frac{\pi}{2}\right) \\
f_2(x) &= (1 + g(x)) \cos\left(x_1 \frac{\pi}{2}\right) \cos\left(x_2 \frac{\pi}{2}\right) \dotsm \sin\left(x_{M-1} \frac{\pi}{2}\right) \\
f_3(x) &= (1 + g(x)) \cos\left(x_1 \frac{\pi}{2}\right) \cos\left(x_2 \frac{\pi}{2}\right) \dotsm \sin\left(x_{M-2} \frac{\pi}{2}\right) \\
& \vdots \\
f_{M-1}(x) &= (1 + g(x)) \cos\left(x_1 \frac{\pi}{2}\right) \sin\left(x_2 \frac{\pi}{2}\right) \\
f_M(x) &= (1 + g(x)) \sin\left(x_1 \frac{\pi}{2}\right)
\end{align*}
Here, $g(x)$ is the sum constraint defined earlier. The solution to this multi-objective function is satisfied when at least $M$ design variables meet the condition $x_i = 0.5$ for $i = 1, 2, \ldots, M$. This function is chosen because, like real-world scenarios, not all design variables influence every objective.
The true Pareto front for the DTLZ-2 problem is known for given $n$ design variables and $M$ objectives. The convergence criterion is based on the computed hyper-volume, with a tolerance of $5\%$ to the true hyper-volume:
\[
\frac{|\text{HV}_{\text{true}} - \text{HV}_i|}{\text{HV}_{\text{true}}} \leq 5\%
\]
To understand the optimization algorithm's complexity and bottlenecks, we study the time and space complexity as functions of $M$ objectives and $n$ design variables. The overall complexity is the convolution of the complexities in building the surrogate model and using the acquisition function. Current studies evaluate these complexities at each step of the Bayesian process. Preliminary results are shown in Fig. \ref{fig:closure-test1}.

Future work will extend this to more combinations of surrogate models and acquisition functions to verify the complexities reported in the literature.


Closure Test 2 aims to efficiently distribute simulations for various optimization tasks through three stages of parallelization. 
First, multi-processing within a single node is used when simulation time is negligible compared to queue time. This is parallelized with Joblib, as in Closure Test 1. Second, parallelization with HPC workload managers like \texttt{SLURM} and PBS is applied when simulation time is comparable to queue time. 
Third, parallelization with PanDA \cite{PanDA-CSBS} ---a data-driven workload management system designed for the LHC data processing scale--- is used when the simulation time is much longer than the queue wait time for job submission. 
%
Closure Test 2 objectives are to integrate the optimization pipeline to use PanDA for evaluating design points and to integrate the \epic software stack with PanDA for detector design evaluation. 
PanDA offers advantages such as OIDC authentication for cross-site submissions and PanDA Monitor for job status tracking. Ongoing efforts include creating authentication pipelines and retrieving information from various sites.


\paragraph{Applications to the ePIC Detector}

We are currently optimizing two detection systems in ePIC: the dual-RICH (dRICH) in the central region and the B0 detector in the forward region.

The dual-radiator Ring Imaging Cherenkov detector (dRICH) will provide charged-hadron particle ID in the forward rapidity region of ePIC for a wide momentum range of \( \sim 2 \, \text{GeV}/c < p < 50 \, \text{GeV}/c \) \cite{VALLARINO2024168834, Cisbani-2020}. The objective of this optimization is to improve pion-kaon and kaon-proton separation, as well as acceptance, at both low and high momentum across the full \( \eta \) range covered by the dRICH.
An initial study applying MOBO to the design of the ePIC dRICH has been conducted, with design points evaluated using the ePIC full simulation. The design parameters used in this optimization are the aerogel disk radius, the spherical mirror position (z,x) and radius (with the z position constrained such that the mirror backplane is within 4 cm of the dRICH back wall), and the photosensor sphere position (z,x) and radius (with the z position constrained such that sensors are within the sensor box required for services). This totals seven design parameters with two hard constraints on the search space. Overlap checks are conducted for each sampled design point before simulation.
For each design point, 1000 \( K^+ \) and 1000 \( \pi^+ \) particles are simulated in six \( p/\eta \) ranges. The momentum values are \( 15 \, \text{GeV}/c \) and \( 40 \, \text{GeV}/c \), at the upper edge of the momentum range covered by each radiator. The \( \eta \) ranges sampled are \([1.3, 2.0], \; [2.0, 2.5], \; [2.5, 3.5]\).
Following \cite{Cisbani-2020}, the objectives chosen for this test optimization are the \( N\sigma \) pion-kaon separation for the low and high \( \eta \) ranges, computed as
\[
N\sigma_{\pi-K} = \frac{|\langle \theta_K \rangle - \langle \theta_{\pi} \rangle| \sqrt{\langle N_{\text{photons}} \rangle}}{\sigma_{\theta}},
\]
where \( \langle \theta \rangle \) is the average reconstructed Cherenkov angle for pions/kaons, \( \langle N_{\text{photons}} \rangle \) is the average number of reconstructed photons per track, and \( \sigma_{\theta} \) is the Cherenkov angle resolution from a Gaussian fit to the Cherenkov angle residual, averaged over pions and kaons. \( N\sigma_{\pi-K} \) is averaged over the two momentum points for each \( \eta \) range.
This optimization was carried out with 50 initial Sobol points, a SAASBO surrogate model, and a qNEHVI acquisition function, with the optimization executed for 200 total evaluations in batches of 10.\footnote{Definitions of Sobol points, SAASBO and qNEHVI can be found in the tutorial \cite{ax_saasbo}.}
Convergence was achieved after 150 trials.  Preliminary Pareto-optimal points found after these trials include potential designs that outperform the nominal dRICH geometry. Initial results shown at the dRICH working group meetings of \epic suggest that while the nominal dRICH geometry is well-optimized for the large \( \eta \) region, the MOBO algorithm identifies opportunities for improvement within the available design space.

Far-forward detectors \cite{FarForward-ProceedingJ} measure very forward neutral and charged particles \cite{EIC-Yellow-Report}. The EIC far-forward detectors consist of four independent subsystems: B0 Tracking and Photon Detection, Roman Pots, Off-Momentum Detectors, and the Zero-Degree Calorimeter. The B0 system measures charged particles in the forward direction and tags neutral particles. Off-momentum detectors measure charged particles from decays and fission, Roman pot detectors measure charged particles near the beam, and the zero-degree calorimeter measures neutral particles at small angles.
The optimization objective for the B0 ECAL calorimeter is to improve momentum resolution under a non-homogeneous magnetic field while simultaneously increasing the B0 ECAL acceptance. Acceptance is defined as the ratio of the number of tracks before the first tracking disk to the number of showers detected by the B0 ECAL. This is challenging because the magnetic field peaks within a limited range in \( z \), and moving it to extremes can disrupt acceptance.
There are four main design parameters, specifically the disk z positions. Some constraints have also been applied to reflect the real scenario, such as all trackers should be placed within 6.65 m, and the sum of the z-positions of two subsequent trackers should be \(\geq 10\) cm.

\begin{figure}[!ht] 
    \centering
    \includegraphics[trim={0cm 0cm 0cm 0cm},clip,width=0.43\textwidth]{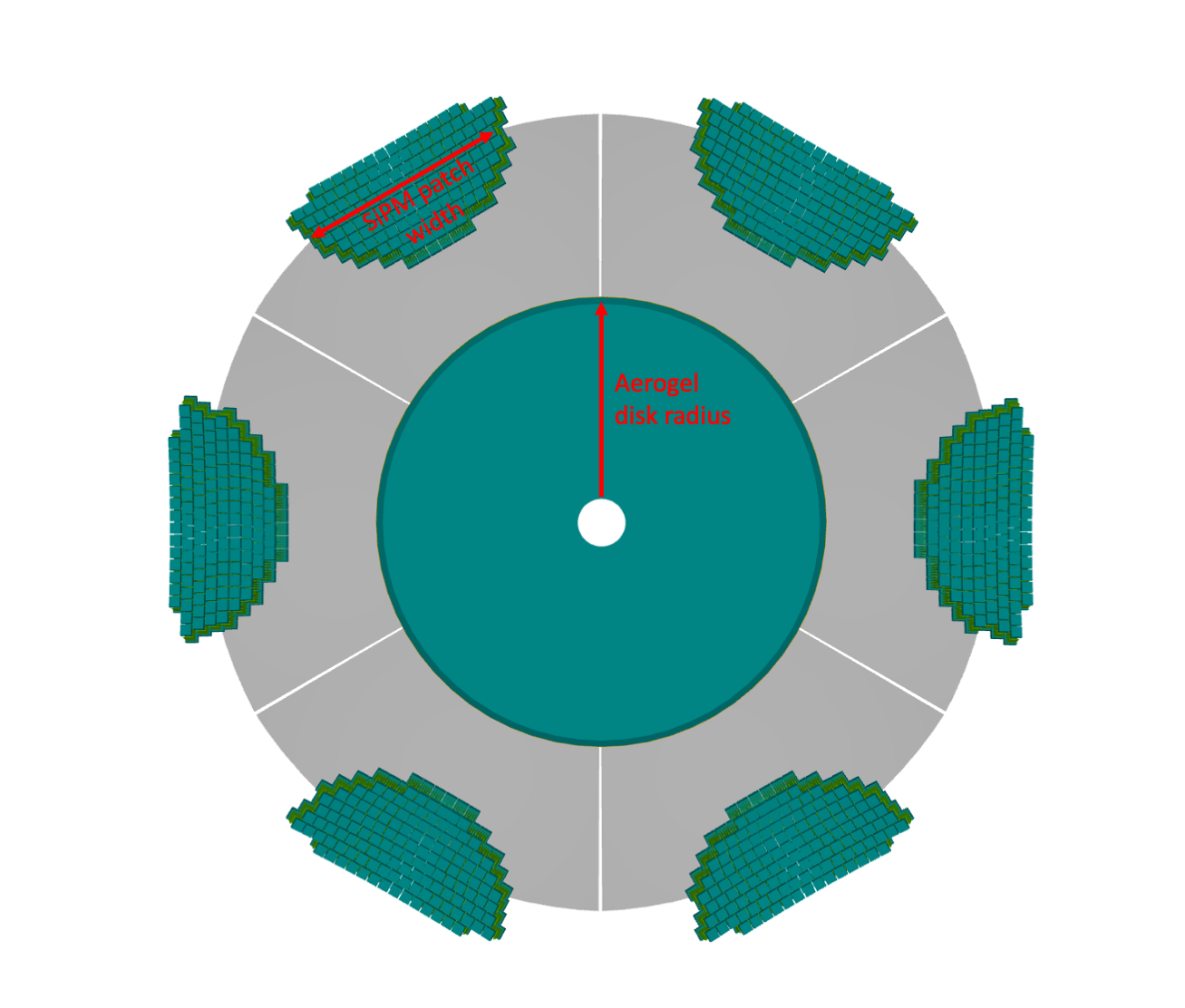}
    \includegraphics[trim={3cm 0cm 3cm 0cm},clip,width=0.55\textwidth]{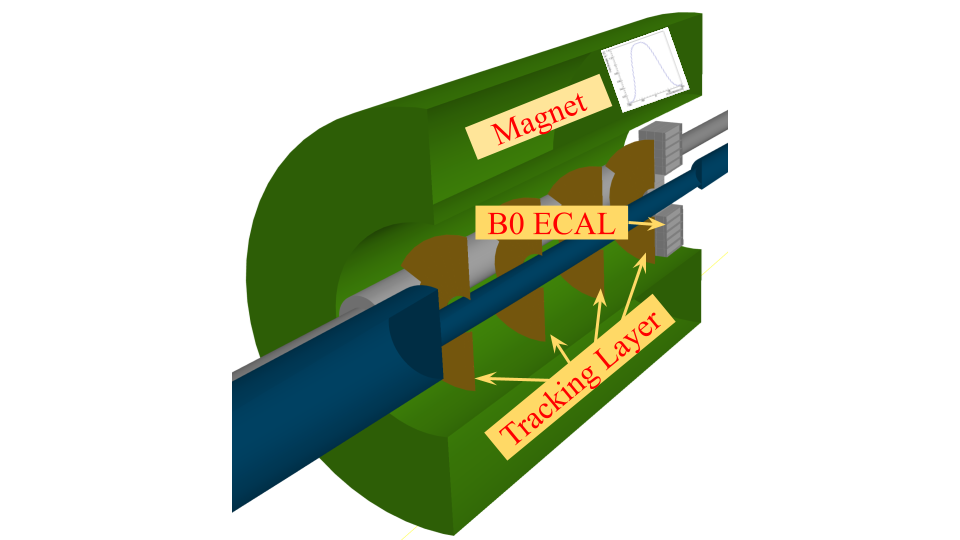}  
    \caption{\textbf{Sub-detector system in \epic optimized with AID2E}: (left) the dRICH design is characterized by two radiators (aerogel and gas) and 6 identical sectors with photosensors \cite{cisbani2020ai}; (right) The B0 Tracker and the B0 ECAL detector. The magnetic field in the B0 Region is inhomogenous making the optimization more challenging.  
    }
    \label{fig:dRICH_B0}
\end{figure}

A visualization of the sub-detecting systems described in this section is shown in Fig. \ref{fig:dRICH_B0}.

%
%

\section{Broader Impact and Future Work}
\label{sec:impact}

Our work with the dRICH and B0 sub-detectors at the \epic experiment demonstrates the potential for extending our methodologies to other sub-detector systems, tailoring to the \epic detector needs. Following the approach implemented in \cite{Fanelli:2022nima}, the tracking system preceding the dRICH is a prime candidate for further optimization in combination with the dRICH. This method is largely experiment-agnostic and applicable to various detection systems, although MOBO currently handles up to four objective functions effectively. For more complex scenarios, the MOGA approach currently available within the AID2E framework offers a viable alternative. 

Large-scale, compute-intensive applications often face resource-based limitations that hinder iterative development, training, and optimization.
Our framework facilitates AI/ML workflows across geographically distributed, multi-platform sites, significantly enhancing the scalability and accessibility for researchers. This reduces barriers from small-scale development to large-scale deployment, enabling deeper and more complex model processing that can yield substantial scientific benefits.
The potential applications of this framework include:
(i) \textbf{Design of future large-scale experiments} in nuclear physics and high-energy physics, such as the early-stage second EIC detector, optimizing for complementarity with the \epic detector across their full physics reach;
(ii) \textbf{Alignments and calibrations}, addressing compute-intensive and multi-objective needs in detector operations;
(iii) \textbf{Multi-objective hyperparameter optimization} for deep neural networks, aiming to balance network performance against complexity, as shown in \cite{efficienct_mobo_nn}.

\section{Documentation and Outreach}

We will employ GitBook along with other knowledge-sharing platforms to support our documentation and outreach efforts, ensuring that these resources are easily accessible to beginners. Automated documentation will be implemented using a Retrieval-Augmented Generation (RAG)-based approach, as described in \cite{suresh2024towards}, with an AI4EIC bot available to the public via a specific URL.
The AID(2)E project employs a comprehensive suite of data science tools to analyze the results of various optimizations. Interested readers can refer to \cite{moo_dashboard} for a dashboard that displays the performance of closure tests, and to \cite{pareto_interactive} for an example of interactive navigation of Pareto front solutions from the ECCE tracker design.
This project will also feature summer bootcamps on AI-assisted design optimization at William \& Mary, incorporating lectures cited in \cite{cfteach-nnps} and \cite{cfteach-hugs}, along with hands-on research activities with the AID(2)E group. Participants will engage in cutting-edge research, learning about AI applications in detector design and data science for physics.
Additional information about the project is available in the main repository, detailed in \cite{aid2e_repository}.

\section{Conclusions}
\label{sec:conclusions}

We are actively working to integrate Multi-Objective Optimization with the EIC shell software, and initial results have already been obtained for two detector systems: the dRICH in the central region and the B0 in the far-forward region of \epic. This development represents a significant milestone, testing the capabilities of our approach in a real-world scenario.
Ultimately, our goal is to develop a scalable and distributed framework capable of holistically optimizing large-scale detectors. Detector-2 at the EIC currently presents an ideal application for this framework, considering its developmental stage.
During construction phases of detector projects, numerous changes are often necessary, such as adjustments due to material availability or budget constraints. The AID(2)E framework is the ideal tool to optimize these design modifications efficiently, utilizing objectives like cost reduction.
Moreover, this framework is poised to have broader impacts beyond its immediate application. It is versatile enough to be adapted for various experimental setups and is particularly well-suited for compute-intensive tasks that require Multi-Objective Optimization, such as calibrations and alignments or optimization of complex machine learning pipelines.


\acknowledgments C.F., K.S. and H.N. were supported by the Office of Nuclear Physics of the U.S. Department of Energy under Grant Contract No. DE-SC0024625, C.P. and K.N. were supported by the Office of Nuclear Physics of the U.S. Department of Energy under Grant Contract No. DE-SC0024478


\bibliography{biblio}

\end{document}